\documentclass[twocolumn]{aastex63} 

\usepackage{graphicx} %Required for including pictures
\usepackage{float} % Allows putting an [H] in \begin{figure} to specify the exact location of the figure
\usepackage{wrapfig} % Allows in-line images such as the example fish picture
\usepackage{lipsum} % Used for inserting dummy 'Lorem ipsum' text into the template
\usepackage{hyperref}
\usepackage{times}
\usepackage{amsmath}
\usepackage{graphicx}
\usepackage{subfigure}
\usepackage{placeins}
\usepackage{hyperref}
\usepackage{gensymb}
\usepackage{upgreek}
\usepackage{natbib}

\usepackage{graphicx}
\usepackage{subfigure}
\usepackage{multirow}
\usepackage{comment}
\usepackage{natbib}
\usepackage{hyperref}
\usepackage{mathtools}
\usepackage{mathrsfs}
\usepackage{fontenc}
\usepackage{color}
\usepackage{url}
\usepackage{hyperref}
\usepackage{gensymb}
\usepackage{pifont}
\graphicspath{{./}}

\newcommand       \K            {\,{\rm K}}
\newcommand       \mum          {{\rm \mu m}}

\submitjournal{ApJ}

\shorttitle{Estimating Molecular Gas Content in Galaxies from PAH Emission}
\shortauthors{Zhang \& Ho}

\begin{document}

\title{Estimating Molecular Gas Content in Galaxies from PAH Emission}

\author[0000-0003-4937-9077]{Lulu Zhang}
\affiliation{Kavli Institute for Astronomy and Astrophysics, Peking University, Beijing 100871, China; l.l.zhang@pku.edu.cn}
\affiliation{Department of Astronomy, School of Physics, Peking University, Beijing 100871, China}

\author[0000-0001-6947-5846]{Luis C. Ho}
\affiliation{Kavli Institute for Astronomy and Astrophysics, Peking University, Beijing 100871, China; l.l.zhang@pku.edu.cn}
\affiliation{Department of Astronomy, School of Physics, Peking University, Beijing 100871, China}

%\email{l.l.zhang@pku.edu.cn}

\begin{abstract} 
Emission from polycyclic aromatic hydrocarbons (PAHs), a commonly used indicator of star formation activity in galaxies, also has the potential to serve as an effective empirical tracer of molecular gas. We use a sample of 19 nearby galaxies with spatially resolved mid-infrared Spitzer spectroscopy, multi-wavelength optical and mid-infrared imaging, and millimeter interferometric CO(1--0) maps to investigate the feasibility of using PAH emission as an empirical proxy to estimate molecular gas mass. PAH emission correlates strongly with CO emission on sub-kpc scales over the diverse environments probed by our sample of star-forming galaxies and low-luminosity active galactic nuclei. The tight observed correlation, likely a consequence of photoelectronic heating of the diffuse interstellar gas by the PAHs, permits us to derive an empirical calibration to estimate molecular gas mass from the luminosity of PAH emission that has a total scatter of only $\sim 0.2-0.25$ dex. Mid-infrared bands sensitive to PAH emission (e.g., the Spitzer/IRAC4 and WISE/W3 filters) can also be used as a highly effective substitute for this purpose.
\end{abstract}


\section{Introduction}\label{sec:Intro}

The formation of stars from dense molecular clouds in the interstellar medium constitutes an essential process in the lifecycle of galaxies. The physical connection between star formation and the interstellar medium can be explored through many observational frameworks, among them the tight correlation between SFR surface density ($\Sigma_{\rm SFR}$) and cold gas surface density ($\Sigma_{\rm gas}$). Originally proposed by \cite{Schmidt 1959} and subsequently extended and refined most notably by \cite{Kennicutt 1998}, the Kennicutt-Schmidt law stipulates that star formation in galaxies conforms to a power-law relation of the form $\Sigma_{\rm SFR} \propto \Sigma_{\rm gas}^{N}$, where $N \approx 1.4$ if both atomic (H~{\small I}) and molecular hydrogen (H$_2$) are included, or $N \approx 1.0$ if only H$_2$ is included (e.g., \citealt{Bigiel et al. 2008, Leroy et al. 2013}). To fully understand the physical processes that regulate galaxy formation and evolution, large galaxy samples with reliable measurements of SFR and gas content are indispensable.

While a broad range of SFR estimators has been extensively developed and applied to normal star-forming galaxies (see reviews in \citealt{Kennicutt & Evans 2012} and \citealt{Madau & Dickinson 2014}) and galaxies hosting active galactic nuclei (AGNs; e.g., \citealt{Zhuang & Ho 2019, Zhuang et al. 2019, Xie et al. 2021}), the measurement of molecular gas content is quite limited owing to the difficulties of direct H$_2$ detection. In practice, most of what we know about the molecular gas in galaxies comes from observations of ``tracer'' species. The most commonly employed tracer of H$_2$ is CO(1--0) emission (e.g., \citealt{Young & Scoville 1991, Solomon & Vanden Bout 2005}; see review in \citealt{Saintonge & Catinella 2022}), but in recent years increasing attention has been devoted to H$_2$ estimators based on dust emission (e.g., \citealt{Magdis et al. 2011, Scoville et al. 2014, Shangguan et al. 2018}) and absorption (\citealt{Concas & Popesso 2019, Yesuf & Ho 2019, Yesuf & Ho 2020, Barrera-Ballesteros et al. 2020}). The CO-based method relies on an uncertain CO-to-H$_2$ conversion factor (\citealt{Blanc et al. 2013, Bolatto et al. 2013}), and a metallicity-dependent dust-to-gas mass ratio needs to be invoked to translate the dust continuum emission to molecular gas mass (\citealt{Galametz et al. 2011, Remy-Ruyer et al. 2014}).

PAH emission, which arises primarily from the warm layer of photodissociation regions in proximity to H~{\small II} regions, is closely associated with both dense, star-forming gas and more diffuse interstellar clouds. Within a photodissociation region, most of the incident ultraviolet radiation from the H~{\small II} region is absorbed by PAH molecules and dust grains of size somewhat larger than the PAHs (\citealt{Bakes & Tielens 1994, Wolfire & Kaufman 2011}). After absorbing an ionizing ultraviolet photon, a PAH molecule mostly undergoes radiative deexcitation through IR emission, which accounts for up to 20\% of the total IR power of galaxies (\citealt{Smith et al. 2007, Diamond-Stanic & Rieke 2010, Xie et al. 2018}). PAH molecules also can deexcite through photodissociation or ionization (\citealt{Allamandola et al. 1989, Allain et al. 1996}), in the process converting $0.1\%–1\%$ of the absorbed energy to energetic photoelectrons that dominate the heating of photodissociation regions as well as the diffuse interstellar clouds (\citealt{Bakes & Tielens 1994, Weingartner & Draine 2001}). These sources of heating in the photodissociation region are balanced by cooling through collisionally excited fine-structure lines of [C~{\small II}]~158~$\mum$ and [O~{\small I}]~63~$\mum$, as well as collisionally excited rotational-vibrational lines of H$_2$ and rotational lines of CO and other abundant molecules (\citealt{Tielens & Hollenbach 1985, Wolfire et al. 1989, Meijerink & Spaans 2005}; see review by \citealt{Hollenbach & Tielens 1997}).

In addition to the above coupling between heating and cooling processes, \cite{Bendo et al. 2008, Bendo et al. 2010, Bendo et al. 2020} proposed, on the basis of the spatial coincidence between PAH and cold dust emission observed within galaxies, that PAH molecules can be mixed to varying degrees with the diffuse cold dust, and hence cold gas. Theoretical calculations and observations indicate that not only ionizing ultraviolet photons from newborn stars but also non-ionizing radiation from evolved stars contribute to the excitation of PAH molecules (e.g., \citealt{Draine & Li 2001, Sellgren 2001, Haas et al. 2002, Li & Draine 2002, Zhang & Ho 2022}). Under such circumstances, the interstellar radiation field simultaneously heats the PAHs and the diffuse interstellar dust, and, so too, the gas. The intimate mixture of PAHs and cold dust ensures a tight physical coupling between PAHs and the gas component in galaxies. Moreover, PAH emission, as an effective indicator of star formation activity in diverse environments (e.g., \citealt{Calzetti et al. 2007, Farrah et al. 2007, Pope et al. 2008, Treyer et al. 2010, Shipley et al. 2016, Maragkoudakis et al. 2018, Xie & Ho 2019}), may be linked especially closely to the cold molecular gas from which new stars form. 

Within this backdrop, we wish to explore the degree to which PAH emission can be used as a proxy to estimate molecular gas content in galaxies. This question is especially timely in light of the enormous improvements in MIR spectroscopy afforded by the successful operation of the James Webb Space Telescope \citep{Rigby et al. 2022}. Indeed, previous studies using the Spitzer Space Telescope \citep{Houck et al. 2004} already have suggested that in nearby galaxies PAH emission correlates with molecular gas (e.g., \citealt{Regan et al. 2006, Pope et al. 2013, Cortzen et al. 2019, Gao et al. 2019, Chown et al. 2021}), although the correlation is non-linear (e.g., \citealt{Regan et al. 2006, Bendo et al. 2008}) and the two components appear to be spatially offset (e.g., \citealt{Casasola et al. 2008, Schinnerer et al. 2013}). These previous investigations were based on broad-band photometry using MIR filters that contain strong PAH features (e.g., IRAC4, which contains PAH~7.7 and 8.6~$\mum$; W3, which contains PAH~8.6, 11.3, and 12.7~$\mum$), which cannot separate the emission features from the underlying continuum, or they relied on spatially integrated spectroscopy with the Spitzer Infrared Spectrograph (IRS; \citealt{Werner et al. 2004}). Our study improves upon previous efforts by taking advantage of spatially resolved analysis of the PAH and molecular gas using a sample of nearby galaxies with available mapping-mode IRS observations as well as interferometric maps of the cold gas (CO and H~{\small I}). We explore the relationship between PAH and CO emission, taking into consideration their mutual correlation with SFR, with the ultimate aim of offering a new, improved empirical predictor of molecular gas content based on the observed strength of PAH emission and star formation activity.

\section{Observational Material}\label{section:sec2}

\subsection{Sample and Data Sets}\label{section:sec2.1}

\begin{figure}[!t]
\center{\includegraphics[width=1\linewidth]{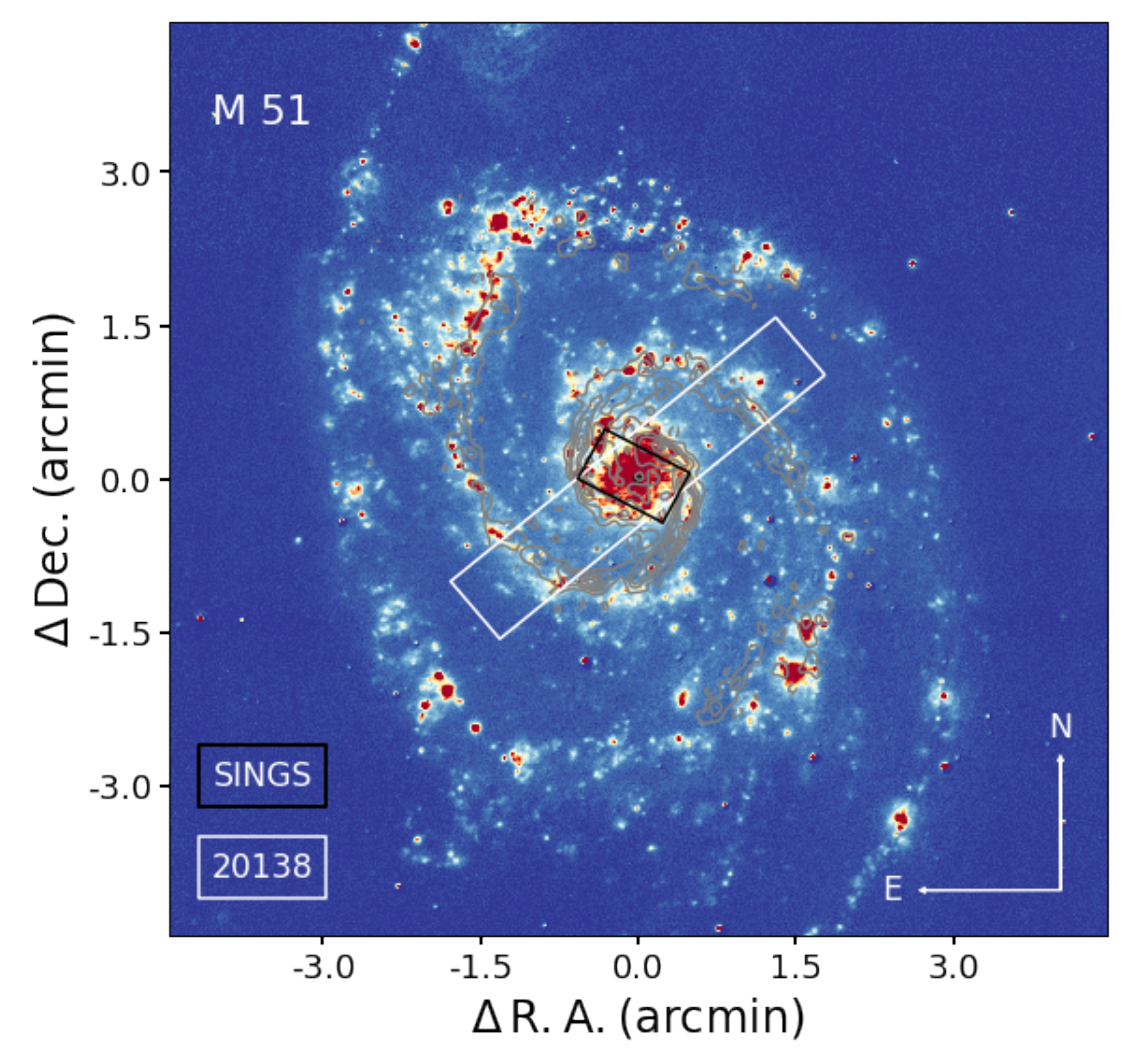}}
\caption{{Illustration of the coverage of IRS mapping-mode observations of M\,51, overlaid on its H$\alpha$ image as the background. The gray contours show the CO distribution. A black star marks the center. The black and white rectangles respectively delineate the coverage of the SINGS observation and that of a separate dedicated observation (ID\,20138) for M\,51.}\label{fig:coverage}}
\end{figure}

The sample is selected from the Spitzer Infrared Nearby Galaxies Survey\footnote{\url{https://irsa.ipac.caltech.edu/data/SPITZER/SINGS/}} (SINGS; \citealt{Kennicutt et al. 2003}), whose mapping-mode IRS observations provide us with the requisite spatially resolved PAH data on 75 nearby star-forming galaxies (SFGs) and galaxies hosting low-luminosity AGNs. SINGS provides a rich repository of multi-wavelength optical and IR images useful for our analysis \citep{SINGS 2020}. The mapping-mode observations of SINGS cover the central $\sim$30\arcsec$\times$50\arcsec\ region of each galaxy over the wavelength range $\sim 5-38\ \mum$ with a spectral resolution of $\lambda/\Delta\lambda\approx 64-128$, depending on the observing module of IRS. The point-spread function (PSF) of IRS increases with wavelength, from full-width at half-maximum (FWHM) $\sim 1\farcs5-3\farcs5$ at $\sim5-14\ \mum$ to ${\rm FWHM} \approx 4\farcs5-9\farcs0$ at $\sim14-38\ \mum$. 

To evaluate the relative effectiveness of estimating PAH strength through MIR broad-band imaging, we analyze images acquired by SINGS using the Spitzer IRAC4 band ($\lambda_{\rm eff}=7.87\,\mum$; \citealt{Fazio et al. 2004}), as well as images taken with the W3 band ($\lambda_{\rm eff}=11.56\,\mum$) as part of the all-sky survey of the Wide-field Infrared Survey Explorer\footnote{\url{https://irsa.ipac.caltech.edu/applications/wise/}} (WISE; \citealt{Wright et al. 2010, IPAC 2020}). We use continuum-subtracted narrow-band H$\alpha$ images, corrected for contamination by [N~{\small II}]~$\lambda\lambda6548, 6584$ in the H$\alpha$ bandpass \citep{Kennicutt et al. 2008}, to trace the distribution of SFR. The IRAC4, W3, and H$\alpha$ images have PSFs with ${\rm FWHM} \approx 2\arcsec, 6\farcs5$, and $2\arcsec$, respectively. The galaxy M\,51 deserves special mention. Although formally contained in SINGS, M\,51 has IRS mapping-mode observations with more extensive spatial coverage ($\sim$40\arcsec$\times$240\arcsec) from a separate, dedicated observations (program 20138; PI: K. Sheth; \citealt{IRSA 2022})\footnote{\url{https://sha.ipac.caltech.edu/applications/Spitzer/SHA/}}. These data were analyzed by \cite{Zhang et al. 2021} and will be used here.

\startlongtable
\begin{deluxetable*}{lccccc||lccccc}
\tabletypesize{\small}
\tablecolumns{12}
\tablecaption{Properties of the Galaxy Sample}
\tablehead{
\colhead{Galaxy} & \colhead{Morphology} & \colhead{Class} & \colhead{{\it D} (Mpc)} & \colhead{{\it i} ($^{\circ}$)} & \colhead{CO/H~{\small I}} & \colhead{Galaxy} & \colhead{Morphology} & \colhead{Class} &\colhead{{\it D} (Mpc)} & \colhead{{\it i} ($^{\circ}$)} & \colhead{CO/H~{\small I}}\\
\colhead{(1)} & \colhead{(2)} & \colhead{(3)} & \colhead{(4)} & \colhead{(5)} & \colhead{(6)} & \colhead{(1)} & \colhead{(2)} & \colhead{(3)} & \colhead{(4)} & \colhead{(5)} & \colhead{(6)}
}
\startdata
NGC~628&SAc&SFG&7.2&24&\ding{52}/\ding{52} & NGC~4450&SAab&AGN&20.0&41&\ding{52}/\ding{56}\\
NGC~925&SABd&SFG&9.12&56&\ding{52}/\ding{52} & NGC~4559&SABcd&SFG&6.98&66&\ding{52}/\ding{56}\\
NGC~2976&SAc&SFG&3.55&63&\ding{52}/\ding{52} & NGC~4579&SABb&AGN&16.4&37&\ding{52}/\ding{56}\\
NGC~3031&SAab&AGN&3.5&58&\ding{52}/\ding{52} & NGC~4725&SABab&AGN&11.9&45&\ding{52}/\ding{56}\\
NGC~3184&SABcd&SFG&11.7&21&\ding{52}/\ding{52} & NGC~4736&SAab&AGN&4.66&35&\ding{52}/\ding{52}\\
NGC~3351&SBb&SFG&9.33&40&\ding{52}/\ding{52} & NGC~4826&SAab&AGN&5.27&54&\ding{52}/\ding{52}\\
NGC~3521&SABbc&AGN&11.2&58&\ding{52}/\ding{52} & NGC\,5194\tablenotemark{\dag}&SABbc&AGN&8.2&15&\ding{52}/\ding{52}\\
NGC~3627&SABb&AGN&9.38&63&\ding{52}/\ding{52} & NGC~6946&SABcd&SFG&6.8&54&\ding{52}/\ding{52}\\
NGC~3938&SAc&SFG&17.9&24&\ding{52}/\ding{56} & NGC~7331&SAb&AGN&14.5&62&\ding{52}/\ding{52}\\
NGC~4321&SABbc&AGN&14.3&32&\ding{52}/\ding{56} & & & & & \\
\enddata
\tablecomments{Col. (1): Galaxy name. Col. (2): Optical morphology from the NASA/IPAC Extragalactic Database. Col. (3): Nuclear classification into star-forming galaxy (SFG) or active galactic nucleus (AGN) according to the [N~{\footnotesize II}]/H$\alpha$ versus [O~{\footnotesize III}]/H$\beta$ emission-line diagnostic diagram based on spectra extracted from a central aperture of 2\farcs5$\times$2\farcs5 (\citealt{Moustakas et al. 2010}). Col. (4): Distance from the updated photometry catalog for SINGS (\citealt{Dale et al. 2017}). Col. (5): Inclination angle (\citealt{Helfer et al. 2003}). Col. (6): Availability of CO and H~{\footnotesize I} maps from the BIMA SONG and THINGS survey, respectively.}
\tablenotetext{\dag}{M\,51 with IRS data cube of much larger coverage; see Section~\ref{section:sec2.1}.}
\label{tab:tablegal}
\end{deluxetable*}

To secure spatially resolved CO maps of comparable resolution, we cross-matched SINGS with the BIMA Survey Of Nearby Galaxies\footnote{\url{https://ned.ipac.caltech.edu/level5/March02/SONG/SONG.html}} (BIMA SONG; \citealt{Helfer et al. 2003, NED 2019}), which was carried out with the 10-element Berkeley-Illinois-Maryland Association (BIMA) millimeter interferometer (\citealt{Welch et al. 1996}). BIMA SONG systematically imaged the 3 mm CO(1--0) emission of the central $\sim$190\arcsec\ of 44 spiral galaxies at an average distance of 12~Mpc, achieving a typical resolution of ${\rm FWHM} \approx 6\arcsec$, which is roughly comparable to that of the PAH data. The combination of SINGS and BIMA SONG yields 19 galaxies with all the required spectral and imaging data sets, which constitute the sample of this study (Table~\ref{tab:tablegal}). We further looked for overlap of the sample galaxies with The HI Nearby Galaxy Survey\footnote{\url{https://www2.mpia-hd.mpg.de/THINGS/Data.html}} (THINGS; \citealt{Walter et al. 2008}), whose Very Large Array H~{\small I} 21~cm maps also have a resolution of ${\rm FWHM} \approx 6\arcsec$, to investigate the possible connection between PAH emission and neutral atomic hydrogen. Of the 34 nearby galaxies covered by THINGS, 13 are in common with our sample.

\subsection{Data Processing and PAH Measurement}\label{section:sec2.2}

\cite{Zhang et al. 2021, Zhang et al. 2022} devised an optimal strategy to extract from low-resolution IRS mapping-mode observation spatially resolved measurements of the integrated ($5 - 20\ \mum$) PAH emission, as well as the strengths of prominent individual PAH features. Where the IRS observations suffered from incomplete wavelength coverage, MIR photometry provided complementary coverage that still enabled effective spectral decomposition. Here we only give a very brief summary of the technical details.

\startlongtable
%\begin{longrotatetable}
%\centerwidetable
%\movetableright=-1in
\begin{deluxetable*}{lccccccc}
\tabletypesize{\small}
\tablecolumns{8}
\tablecaption{Spatially Resolved Measurements of Surface Brightness of PAH and Gas Components}
\tablehead{
\colhead{Region} & \colhead{$A$} & \colhead{log $\Sigma_{\rm PAH}$} & \colhead{log $\Sigma_{\rm CO}$} & \colhead{log $\Sigma_{\rm H~I}$} & \colhead{log $\Sigma_{\rm H\alpha}$} & \colhead{log $\Sigma_{\rm IRAC4}$} & \colhead{log $\Sigma_{\rm W3}$} \\
\colhead{} & \colhead{($\rm pc^{2}$)} & \colhead{($\rm erg\ s^{-1}\ pc^{-2}$)} & \colhead{($\rm erg\ s^{-1}\ pc^{-2}$)} & \colhead{($\rm erg\ s^{-1}\ pc^{-2}$)} & \colhead{($\rm erg\ s^{-1}\ pc^{-2}$)} & \colhead{($\rm erg\ s^{-1}\ pc^{-2}$)} & \colhead{($\rm erg\ s^{-1}\ pc^{-2}$)}\\
\colhead{(1)} & \colhead{(2)} & \colhead{(3)} & \colhead{(4)} & \colhead{(5)} & \colhead{(6)} & \colhead{(7)} & \colhead{(8)}}
\startdata
NGC0628\_1 & $2.44\times10^5$ & $34.85 \pm 0.09$ & $30.23 \pm 0.07$ & $25.26 \pm 0.95$ & $33.46 \pm 0.02$ & $35.20 \pm 0.04$ & $34.97 \pm 0.01$ \\
NGC0628\_2 & $2.44\times10^5$ & $34.89 \pm 0.09$ & $30.15 \pm 0.08$ & $25.37 \pm 0.74$ & $33.36 \pm 0.03$ & $35.12 \pm 0.04$ & $34.97 \pm 0.01$ \\
NGC0628\_3 & $3.66\times10^5$ & $34.96 \pm 0.09$ & $30.11 \pm 0.07$ & $25.34 \pm 0.63$ & $33.31 \pm 0.02$ & $35.10 \pm 0.03$ & $34.92 \pm 0.01$ \\
NGC0628\_4 & $4.87\times10^5$ & $34.88 \pm 0.09$ & $30.11 \pm 0.06$ & $25.53 \pm 0.36$ & $33.24 \pm 0.02$ & $35.02 \pm 0.03$ & $34.83 \pm 0.01$ \\
NGC0628\_5 & $1.22\times10^5$ & $34.78 \pm 0.09$ & $30.19 \pm 0.10$ & $25.46 \pm 0.84$ & $33.19 \pm 0.05$ & $35.04 \pm 0.05$ & $34.83 \pm 0.02$ \\
NGC0925\_1 & $7.82\times10^5$ & $34.60 \pm 0.09$ & \nodata          & $26.34 \pm 0.09$ & $33.38 \pm 0.03$ & $34.88 \pm 0.03$ & $34.58 \pm 0.01$ \\
NGC0925\_2 & $2.15\times10^6$ & $34.43 \pm 0.09$ & \nodata          & $26.21 \pm 0.08$ & $33.21 \pm 0.02$ & $34.60 \pm 0.02$ & $34.37 \pm 0.01$ \\
NGC2976\_1 & $5.92\times10^4$ & $34.79 \pm 0.09$ & $29.65 \pm 0.17$ & $26.14 \pm 0.12$ & $33.61 \pm 0.04$ & $34.94 \pm 0.03$ & $34.73 \pm 0.01$ \\
NGC2976\_2 & $5.92\times10^4$ & $34.75 \pm 0.09$ & $29.23 \pm 0.37$ & $26.18 \pm 0.11$ & $33.31 \pm 0.07$ & $34.87 \pm 0.03$ & $34.64 \pm 0.01$ \\
NGC2976\_3 & $5.92\times10^4$ & $34.68 \pm 0.09$ & $29.45 \pm 0.23$ & $26.05 \pm 0.14$ & $33.51 \pm 0.05$ & $34.85 \pm 0.03$ & $34.63 \pm 0.01$ \\
\enddata
\tablecomments{Col. (1): Resolved spaxel. Col. (2): Spaxel area. The surface brightness and associated uncertainty for the integrated PAH emission of the best-fit PAH template from $5 - 20\ \mum$ (Col. 3), CO(1--0) emission (Col. 4), H~{\small I} emission (Col. 5), extinction-corrected H$\alpha$ emission (Col. 6), IRAC4-band emission, corrected for the stellar contribution based on the IRAC1 map with a scaling factor of 0.232 (following \citealt{Helou et al. 2004}).  (Col. 7), and W3-band emission (Col. 8). {\it (This table is available in its entirety in machine-readable format. A portion is shown here for guidance regarding its form and content.)}}
\label{tab:TableGasCalSigma}
\end{deluxetable*}
%\end{longrotatetable}

After fitting and subtracting the residual background with a two-dimensional polynomial function, all the ancillary MIR and H$\alpha$ images were convolved to a common angular resolution, which is constrained by the coarsest PSF of the IRS spectral data cube (${\rm FWHM} \approx 9\farcs0$). We matched the resolution of the CO and H~{\small I} maps in a similar manner, using the {\tt Gaussian2D} function in {\tt astropy.modeling.models} (\citealt{Astropy Collaboration et al. 2013}) to mimic their original elliptical synthesized beam and the {\tt create\_matching\_kernel} function in {\tt photutils}\footnote{\url{https://media.readthedocs.org/pdf/photutils/stable/photutils.pdf}} to construct the convolution kernels. The convolution was performed using the {\tt convolve\_fft} function in {\tt astropy.convolution}. After convolving all the data to the same angular resolution, we reprojected them into the same coordinate frame as the PAH map to facilitate the subsequent pixel-based, spatially resolved analysis. The final pixel size of the reprojected data was set to 10\arcsec\ to ensure that all the pixels within each galaxy are spatially independent; this size corresponds to a physical scale of $\sim 0.2 - 0.7$~kpc for the distances of our sample (Table~\ref{tab:tablegal}). Some pixels might require additional optimal binning to achieve the minimal signal-to-noise ratio required for robust PAH measurements \citep{Zhang et al. 2021}. The final number of spatially resolved spaxels for each galaxy ranged from 1 to 15, with the exception of M\,51, which is $\sim$ 100.

The PAH strength was measured using the template-fitting technique of \cite{Xie et al. 2018} from complete (SL+LL) low-resolution IRS spectra. After masking out the isolated ionic emission lines, the IRS spectrum is fit with a multicomponent model consisting of a theoretical PAH template and three continuum components represented by modified blackbodies of different temperatures, all subject to attenuation by foreground dust extinction. The fitting is carried out with the Bayesian Markov chain Monte Carlo procedure {\tt emcee} in the {\tt Python} package, as described in Section~4.2 of \cite{Shangguan et al. 2018}. The median and standard deviation of the posterior distribution of each best-fit parameter are taken as the final estimate and corresponding uncertainty. Based on the best-fit model for each spaxel, we define the total PAH luminosity, $L_{\rm PAH}$, as the integral of the best-fit PAH template over the $5 - 20\ \mum$ region, following the convention of \cite{Zhang et al. 2021}. \cite{Xie et al. 2018} showed that the measurements of individual PAH features based on the method used here are well consistent with measurements from other methods. The template-fitting method obtains $\sim 30\%$ more total PAH flux compared to the PAHFIT \citep{Smith et al. 2007} and CAFE \citep{Marshall et al. 2007} methods because it better recovers the long-wavelength continuum associated with PAH skeletal vibrations.

\section{Results}\label{sec:sec3}
\subsection{The Correlation between Molecular Gas and PAHs}\label{sec:sec3.1}

Previous work on PAH emission as a proxy for molecular gas was based on either global PAH measurements of galaxies (e.g., \citealt{Pope et al. 2013, Cortzen et al. 2019, Gao et al. 2019}) or spatially resolved measurements that use MIR broad-band photometry, which is sensitive to but does not give precise measurement of PAH emission (e.g., \citealt{Regan et al. 2006, Bendo et al. 2010, Chown et al. 2021}). Here we investigate, for the first time, whether PAH emission can trace molecular gas using accurate PAH measurements derived from spectral decomposition of spatially resolved MIR spectra (Section~\ref{section:sec2}). We then used these improved mesurements to attempt to delineate the underlying mechanism for the tight correlation between PAH emission and molecular gas.

Following \cite{Solomon et al. 1997}, we calculate the CO(1--0) luminosity, in units of $\rm erg\ s^{-1}$, as

\begin{align}\label{equ:LumCO}
L_{\rm CO} = 4\pi D_L^2 \times 10^{-23} \nu_{\rm obs}\times S_{\rm CO}\Delta V / c,
\end{align}

\noindent
where $D_{L}$ is the luminosity distance in $\rm cm$, $\nu_{\rm obs} = \nu_{\rm rest} / (1+z)$ is the observed CO(1--0) line frequency, where $\nu_{\rm rest} = 1.153\times10^{11}\,{\rm Hz}$ and $z$ is the redshift, $c$ is the speed of light in $\rm km\ s^{-1}$, and $S_{\rm CO}\Delta V$ is the velocity-integrated flux in $\rm Jy\ km\ s^{-1}$, which is converted from the reprojected moment 0 map in $\rm Jy\ km\ s^{-1}\ beam^{-1}$ by multiplying by $\rm \frac{4ln2\theta^{2}}{\pi FWHM^{2}}$, with $\rm \theta$ the pixel size of the projected map and $\rm FWHM$ the beam size of the original map. The associated uncertainty is

\begin{align}\label{equ:LumDelCO}
\Delta L_{\rm CO} = 4\pi D_L^2 \times 10^{-23} \nu_{\rm obs}\times {\rm rms}\ \Delta v\sqrt{N} / c,
\end{align}

\noindent
with the root-mean-square (rms) noise of the channel map in $\rm Jy$, $\Delta v$ the channel width in $\rm km\ s^{-1}$, and $N$ the number of channels over which the line is integrated \citep{Helfer et al. 2003}. Note that only the spaxels with $\Delta \log {L_{\rm CO}} < 1$ dex are used in this paper. The surface brightness of the CO emission is defined as $\Sigma_{\rm CO} = L_{\rm CO}\,{{\rm cos}~i}/A$, where $i$ is the galaxy inclination angle and $A$ is the spaxel area in $\rm pc^{2}$. Similar definitions apply to H~{\small I}.

We correct the H$\alpha$ emission for Galactic extinction using the extinction curve of \cite{Cardelli et al. 1989} with $R_V = 3.1$, for which $A_{\rm H\alpha} = 2.535 E(B-V)$ (\citealt{Kennicutt et al. 2007}), using $E(B-V)$ from \cite{Schlafly & Finkbeiner 2011}. We account for internal extinction following the prescription of \cite{Calzetti et al. 2007} for resolved H~{\small II} regions, $L_{\rm H\alpha} = L_{\rm H\alpha}^{\rm obs} + 0.031\,L_{\rm 24\,\mum}$, where $L_{\rm H\alpha}^{\rm obs}$ is the Galactic extinction-corrected H$\alpha$ luminosity and $L_{\rm 24\,\mum}$ is derived from MIPS24 images. Then, we adopt the prescription ${\rm SFR} = 5.3\times\,10^{-42}\,L_{\rm H\alpha}$ \citep{Kennicutt & Evans 2012}, assuming implicitly that it applies to galaxies on small ($\lesssim 1$~kpc) scales. As discussed by \cite{Kennicutt & Evans 2012}, statistical approximations used to estimate SFRs may break down on small scales, but it is beyond the scope of this work to address this issue.

\begin{figure*}[!ht]
\center{\includegraphics[width=1\textwidth]{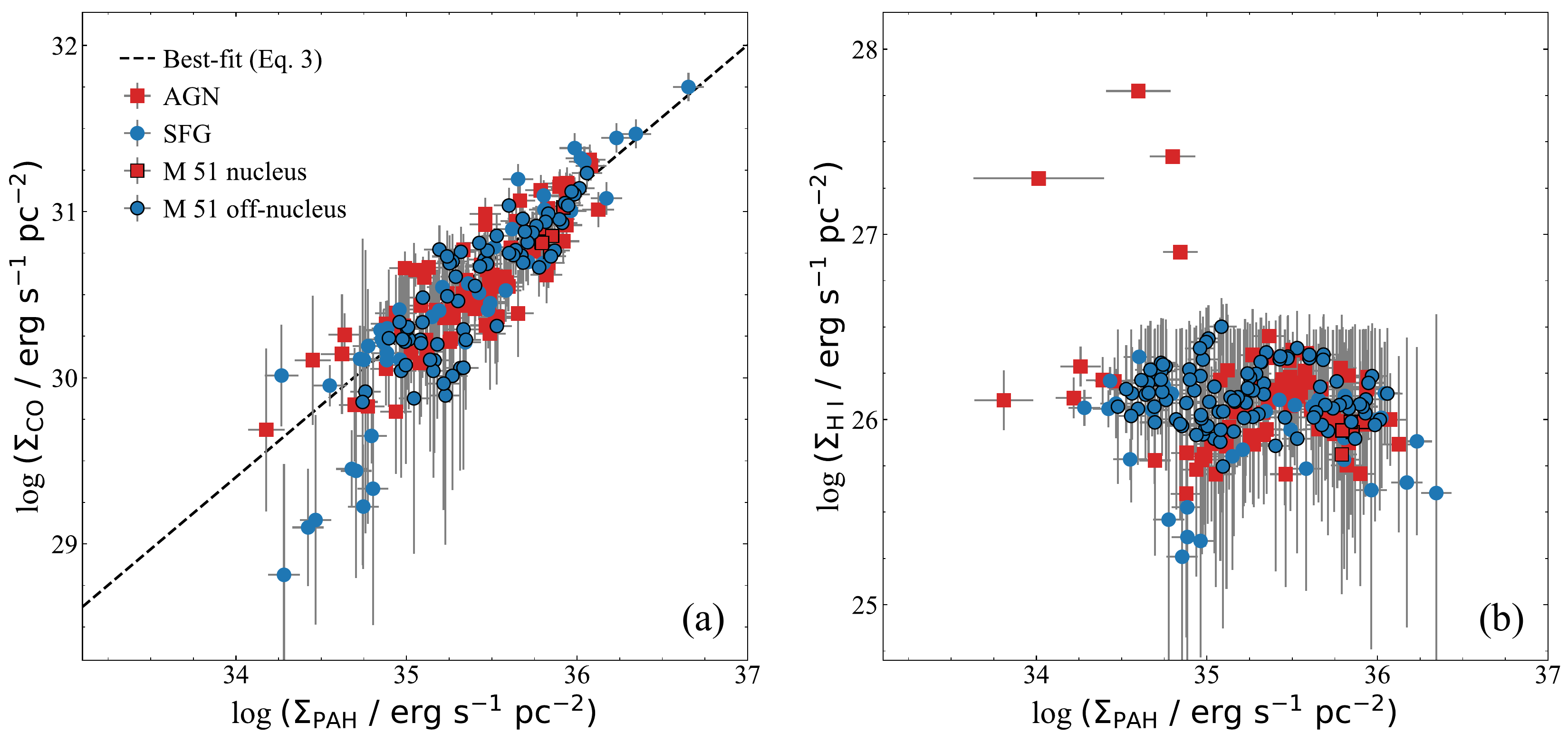}}
\caption{{The correlation between the surface brightness of integrated PAH emission and (a) CO(1--0) and (b) H~{\small I} emission. Red squares and blue circles correspond to spaxels from the central regions of AGNs and SFGs, respectively, while the corresponding symbols with black borders denote the nuclear (AGN-like) and off-nuclear (SFG-like) spaxels from M\,51. The black-dashed line in panel (a) gives the best-fit linear regression (Eq.~\ref{equ:PAH_CO}).}\label{fig:PAH_Gas}}
\end{figure*}

Figure~\ref{fig:PAH_Gas}a examines the relationship between the surface brightness of PAH emission and that of CO(1--0) emission. In light of the evidence that AGNs can adversely affect PAHs (e.g., \citealt{Diamond-Stanic & Rieke 2010, Micelotta et al. 2010a, Micelotta et al. 2010b, Xie & Ho 2022, Zhang et al. 2022}), we distinguish the spaxels from the AGNs and the nuclear region of M\,51 (red points) from those of SFGs and the off-nuclear regions of M\,51 (blue points). Both sets of spaxels show a very strong correlation, and there are no significant differences between them. Thus, for spatially resolved regions of galaxies of different activity types, as well as across different environments within a single galaxy, PAH emission tightly correlates with the molecular gas as traced by CO emission. A linear regression analysis with the {\tt Python} package {\tt linmix} (\citealt{Kelly 2007}) yields

\begin{align}\label{equ:PAH_CO}
\begin{aligned}
&{\rm log}\,\Sigma_{\rm CO} = \\&(0.867\pm0.035)({\rm log}\, \Sigma_{\rm PAH} - 34.0) + (29.402\pm0.056),
\end{aligned}
\end{align}

\noindent
with a total scatter of $\epsilon_t = 0.239$\,dex and an intrinsic scatter of $\epsilon_i = 0.081$\,dex. The nearly linear correlation between CO and integrated PAH emission confirms the feasibility of using PAHs as a proxy for estimating molecular gas, even on sub-kpc scales. For completeness, Appendix~\ref{sec:AppA} presents the results of substituting the spectroscopically derived integrated PAH emission with photometric estimates based on IRAC4 and W3 images.

By sharp contrast, the H~{\small I} line shows no relation with PAH emission whatsoever (Figure~\ref{fig:PAH_Gas}b). The lack of a connection between atomic hydrogen gas and PAHs may not be unexpected, given that the radial profile of H~{\small I} differs from that of molecular hydrogen (e.g., \citealt{Bigiel & Blitz 2012, Wang et al. 2014, Casasola et al. 2017}). Spatially resolved studies of the Large Magellanic Cloud also reveal that CO correlates poorly with H~{\small I} \citep{Wong et al. 2009}. 

\begin{figure}[!t]
\center{\includegraphics[width=1\linewidth]{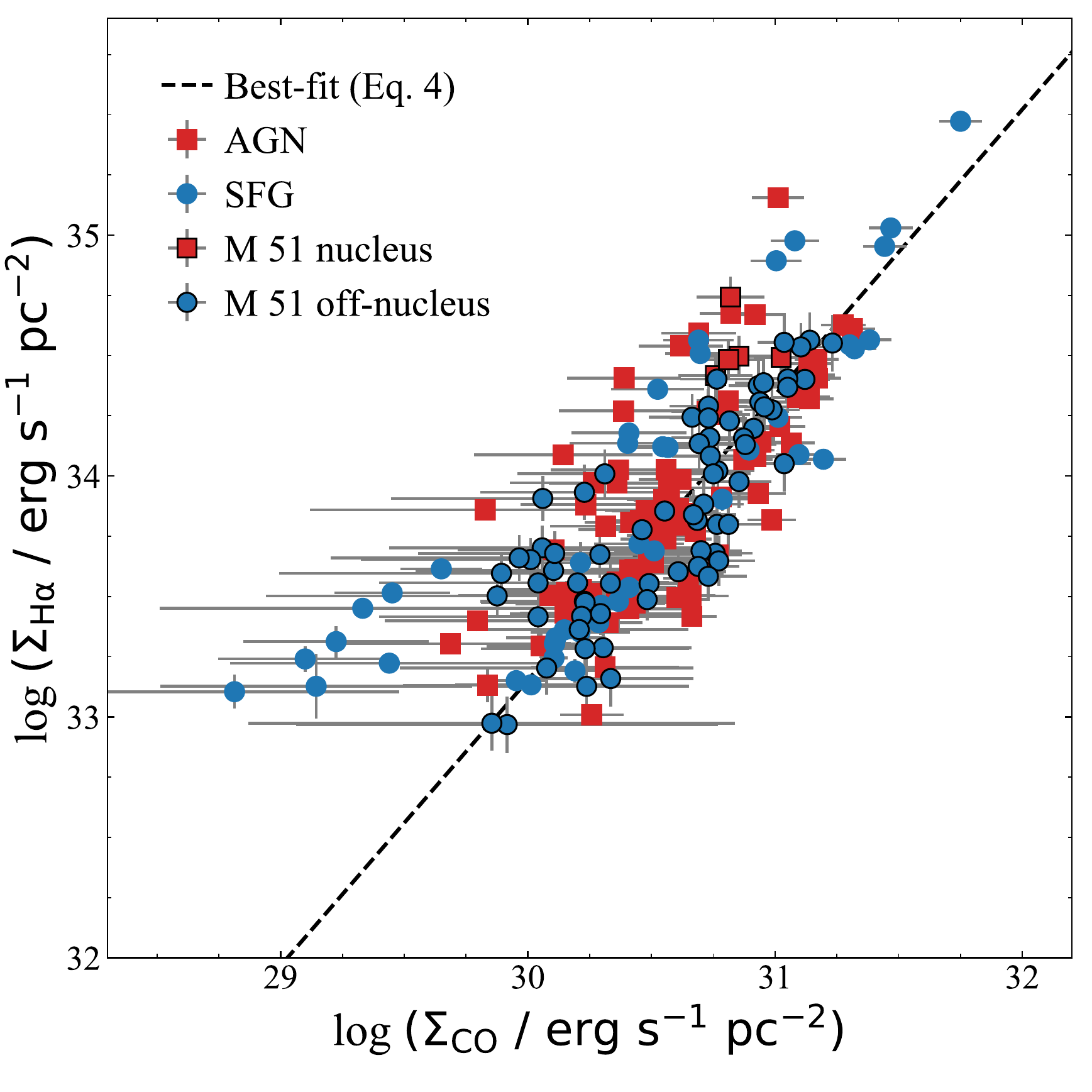}}
\caption{{The correlation between the surface brightness of extinction-corrected H$\alpha$ emission and CO(1--0) emission. Red squares and blue circles correspond to spaxels from the central regions of AGNs and SFGs, respectively, while the corresponding symbols with black borders denote the nuclear (AGN-like) and off-nuclear (SFG-like) spaxels from M\,51. The black-dashed line indicates the best-fit linear regression (Eq.~\ref{equ:Ha_CO}).}\label{fig:Ha_CO}}
\end{figure}

\begin{figure}[!t]
\center{\includegraphics[width=1\linewidth]{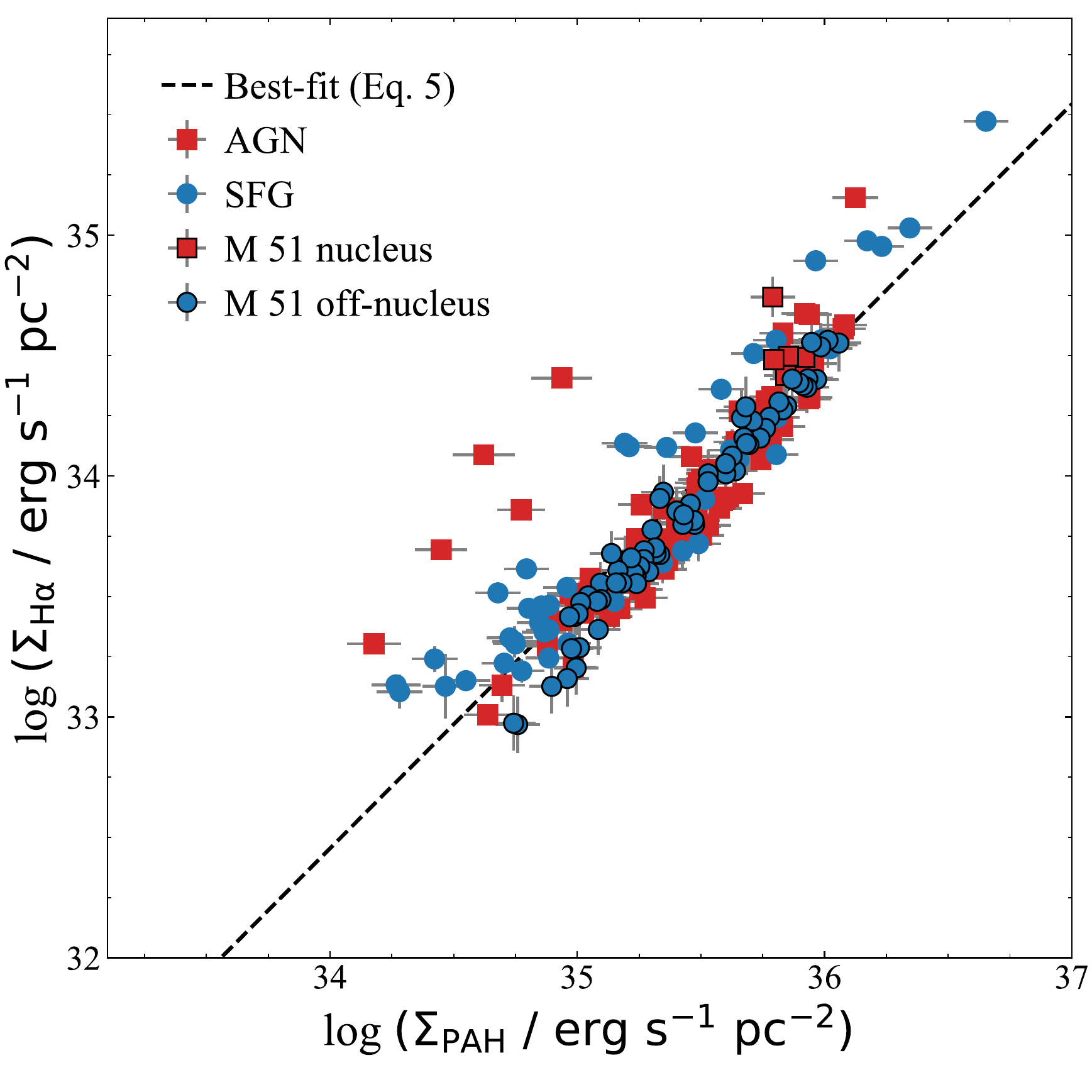}}
\caption{{The correlation between the surface brightness of extinction-corrected H$\alpha$ emission and PAH emission. Red squares and blue circles correspond to spaxels from the central regions of AGNs and SFGs, respectively, while the corresponding symbols with black borders denote the nuclear (AGN-like) and off-nuclear (SFG-like) spaxels from M\,51. The black-dashed line indicates the best-fit linear regression (Eq.~\ref{equ:PAH_Ha}).}\label{fig:PAH_Ha}}
\end{figure}

What is the underlying physical mechanism behind the tight correlation between CO and PAH emission? Could it stem simply from their mutual connection with SFR? After all, SFR empirically correlates with PAH emission, and, at the same time, SFR is physically tied to the molecular gas via the Kennicutt-Schmidt law. Or, perhaps, the CO-PAH correlation reflects a common excitation mechanism (Section~\ref{sec:Intro}). Figure~\ref{fig:Ha_CO} shows the relationship between the surface brightness of the extinction-corrected H$\alpha$ emission and that of the CO emission. As anticipated, the data confirm that the spatially resolved spaxels of the SINGS galaxies obey the Kennicutt-Schmidt relation. A linear regression gives

\begin{align}\label{equ:Ha_CO}
\begin{aligned}
&{\rm log}\,\Sigma_{\rm H\alpha} = \\&(1.185\pm0.067)({\rm log}\, \Sigma_{\rm CO} - 30.0) + (33.152\pm0.053),
\end{aligned}
\end{align}

\noindent
with $\epsilon_t = 0.362$\,dex and $\epsilon_i = 0.220$\,dex. The slope of Equation~\ref{equ:Ha_CO} is consistent with a power-law index of $N\approx1$ for the Kennicutt-Schmidt relation involving only molecular gas, as obtained in previous studies of nearby galaxies spatially resolved on comparable scales (e.g., \citealt{Bigiel et al. 2008, Leroy et al. 2013}). At the same time, as is well-known, H$\alpha$ emission, used here to indicate SFR, also correlates strongly with PAH emission (Figure~\ref{fig:PAH_Ha}; $\epsilon_t = 0.208$\,dex, $\epsilon_i = 0.175$\,dex):

\begin{align}\label{equ:PAH_Ha}
\begin{aligned}
&{\rm log}\,\Sigma_{\rm H\alpha} = \\&(1.030\pm0.036)({\rm log}\, \Sigma_{\rm PAH} - 34.0) + (32.454\pm0.053).
\end{aligned}
\end{align}

The closely similar slopes of Equations~\ref{equ:PAH_CO}--\ref{equ:PAH_Ha} do not offer much guidance as to whether the CO-PAH correlation is physical or merely an artifact of the SFR-CO and SFR-PAH relations. Nevertheless, two arguments can be advanced to suggest that molecular gas does have a direct link with PAHs. On the one hand, among the three empirical relations presented above, that involving CO and PAH (Equation~\ref{equ:PAH_CO}) by far has the lowest formal intrinsic scatter (0.08\,dex). We should be cautious, however, not to overinterpret this result, because the total scatter of all three correlations is likely dominated by many sources of systematic uncertainty, which are difficult to quantify. The exact scatter also depends on the statistical method used for the linear regression. On the other hand, the Spearman correlation between $\Sigma_{\rm CO}$ and $\Sigma_{\rm PAH}$ is highly significant, for all galaxies combined or for SFGs and AGNs separately (correlation coefficient $\rho \approx 0.80-0.90$; $p$-value $< 10^{-20}$). Moreover, it still remains highly significant after taking into consideration the mutual correlation of these two variables with $\Sigma_{\rm H\alpha}$; the partial rank correlation between $\Sigma_{\rm CO}$ and $\Sigma_{\rm PAH}$ while holding $\Sigma_{\rm H\alpha}$ fixed drops only slightly ($\rho \approx 0.60$; $p$-value $< 10^{-15}$). We suggest that the tight correlation between PAH and CO emission is physical and not an artifact of their mutual correlation with SFR. As mentioned in the Introduction, the photoelectronic effect of PAH molecules contributes to the heating of the gas \citep{Wolfire et al. 1989, Stacey et al. 1991}, and this might be the physical basis for the CO-PAH correlation. And while PAHs traditionally are viewed as concentrated toward photodissociation regions, they, in actuality, can be excited by the radiation field of the total stellar population and have a more extended spatial distribution, further enhancing their connection to the broader cold gas component (e.g., \citealt{Jones et al. 2015}; \citealt{Bendo et al. 2020}).

\begin{figure}[!ht]
\center{\includegraphics[width=1\linewidth]{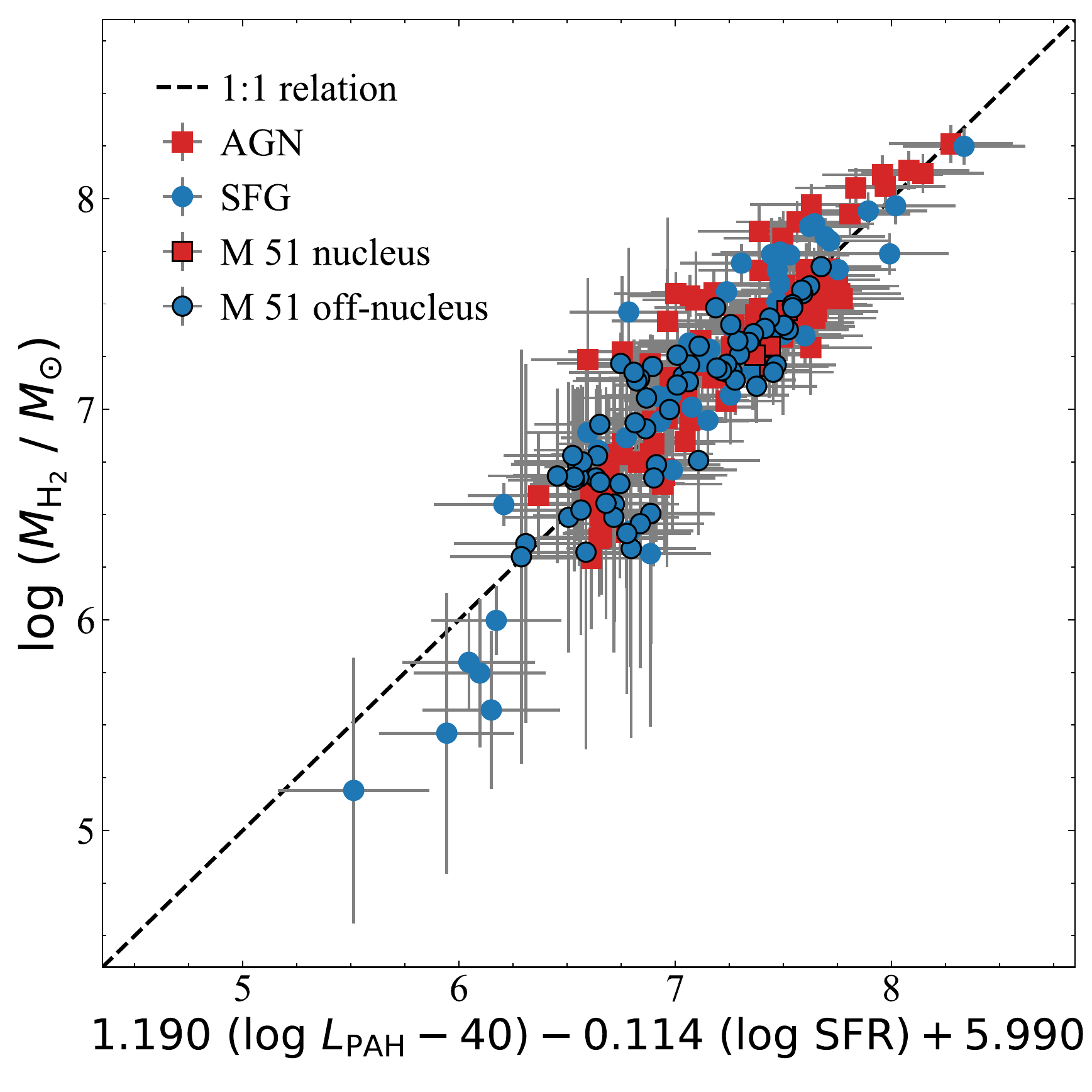}}
\caption{{Estimating molecular gas mass ($M_{\rm {H_2}}$) from PAH luminosity and SFR (Eq.~\ref{equ:H2_PAH_SFR}). Red squares and blue circles correspond to spaxels from the central regions of AGNs and SFGs, respectively, while the corresponding symbols with black borders denote the nuclear (AGN-like) and off-nuclear (SFG-like) spaxels from M\,51. The black-dashed line indicates the 1:1 relation.}
\label{fig:MH2}}
\end{figure}

To explore whether CO can be predicted more accurately by simultaneously combining PAH emission and SFR, we performed multiple linear regression analysis with the {\tt Python} package {\tt statsmodels} (\citealt{Seabold & Perktold 2010}), which yields the following joint fit:

\begin{align}\label{equ:CO_PAH_Ha}
\begin{aligned}
&{\rm log}\,\Sigma_{\rm CO} = (0.877\pm0.081)({\rm log}\, \Sigma_{\rm PAH} - 34.0)\ + \\
&(0.003\pm0.075)({\rm log}\, \Sigma_{\rm H\alpha} - 33.0) + (29.320\pm0.063).
\end{aligned}
\end{align}

\normalsize\noindent
Compared to Equation~\ref{equ:PAH_CO}, Equation~\ref{equ:CO_PAH_Ha} gives only a slightly reduced scatter ($\epsilon_t = 0.226$\ dex, $\epsilon_i = 0.077$\ dex).

\subsection{Estimating Molecular Gas Content from PAH Emission}\label{sec:sec3.2}

Having argued that CO(1--0) emission is strongly correlated with PAH emission, and that their correlation may have a physical basis, we propose an empirical prescription to estimate molecular gas mass from the observed strength of PAH emission. We offer two prescriptions, one that relies solely on the strength of the PAH emission, and another that further introduces the SFR as a secondary variable. In this work, we use the extinction-corrected H$\alpha$ emission to estimate SFR, but other estimators can be substituted. Adopting a canonical CO-to-H$_2$ conversion factor of ${\rm \alpha_{CO}} = 3.2\ M_{\odot}\ {\rm pc^{-2}\ (K\ km\ s^{-1})^{-1}}$ (this value should be multiplied by a factor of 1.36 to include helium; \citealt{Bolatto et al. 2013}), and converting $\Sigma_{\rm CO}$ ($\rm erg\ s^{-1}\ pc^{-2}$) to velocity-integrated antenna temperature $T_{\rm int}$ ($\rm K\ km\ s^{-1}$) using $T_{\rm int} = 10^{3}\frac{c^{3}}{2k\nu_{\rm rest}^{3}}\frac{\Sigma_{\rm CO}}{4\pi\ {\rm pc}^{2}}$, where $k$ is Boltzmann's constant and pc is one parsec unit, we obtain, analogous to Equation~\ref{equ:PAH_CO},

\begin{align}\label{equ:H2_PAH}
\begin{aligned}
&\log\,M_{\rm H_2} = \\&(1.029\pm0.040)(\log\, L_{\rm PAH} - 40.0) + (6.399\pm0.040),
\end{aligned}
\end{align}

\noindent
where $M_{\rm H_2}$ is in units of $M_{\odot}$ and $L_{\rm PAH}$ is in units of $\rm erg\ s^{-1}$. The scatters are $\epsilon_t = 0.231$\ dex and $\epsilon_i = 0.093$\ dex. Including the SFR (in units of $M_{\odot}\rm \ yr^{-1}$) as in Equation~\ref{equ:CO_PAH_Ha}, we have

\begin{align}\label{equ:H2_PAH_SFR}
\begin{aligned}
&{\rm log}\,M_{\rm H_{2}} = \\&(1.190\pm0.089)({\rm log}\, L_{\rm PAH} - 40.0)\ + \\&(-0.114\pm0.079)({\rm log}\,{\rm SFR}) + (5.990\pm0.224),\,\,\,\,\,\,\,\,\,\,\,\,\,\,\,\,\,\,\,\,
\end{aligned}
\end{align}

\noindent
for which $\epsilon_t = 0.226$\ dex and $\epsilon_i = 0.080$\ dex (Figure~\ref{fig:MH2}).  

\begin{table*}
%\begin{center}
\caption{Best-fit Parameters for the $M_{\rm H_2}$ Calibration using Different PAH Bands}
\label{tab:calibration}
\begin{tabular}{ccccccccccc}
\hline
\hline
Band              && $\alpha_L$        && $\alpha_S$         && $\beta$           && $\epsilon_t$ && $\epsilon_i$ \\
\hline
PAH~$5-20\,\mum$  && $1.029\pm0.040$ && \nodata            && $6.399\pm0.040$ && 0.231         && 0.093         \\
                  && $1.190\pm0.089$ && $-0.114\pm0.079$ && $5.990\pm0.224$ && 0.226         && 0.080         \\
\\
IRAC4             && $1.017\pm0.037$ && \nodata            && $6.261\pm0.043$ && 0.225         && 0.111         \\
                  && $1.558 \pm 0.102$ && $-0.465 \pm 0.091$ && $4.783 \pm 0.272$ && 0.203         && 0.059         \\
\\
W3                && $1.061 \pm 0.041$ && \nodata            && $6.397 \pm 0.040$ && 0.238         && 0.130        \\
                  && $1.429 \pm 0.120$ && $-0.300 \pm 0.104$ && $5.460 \pm 0.296$ && 0.224         && 0.122         \\
\hline
\hline
\end{tabular}
%\end{center}
\tablecomments{\footnotesize{Best-fit parameters for $\log M_{\rm H_2}=\alpha_L\, (\log L_{\rm band} - 40.0)\,+\, \alpha_S\, \log {\rm SFR} + \beta$, with $M_{\rm H_2}$ in units of $M_{\odot}$ and $L_{\rm band}$ in units of erg~s$^{-1}$. The total and intrinsic scatter, $\epsilon_t$ and $\epsilon_i$, are in units of dex.}}
\normalsize
\end{table*}

Appendix~\ref{sec:AppA} demonstrates that in the absence of spectroscopy even broad-band photometry using the Spitzer/IRAC4 and WISE/W3 filters can provide quite accurate predictors of CO emission. Table~\ref{tab:calibration} summarizes the best-fit parameters for the $M_{\rm H_{2}}$ calibration based on different measures of PAH strength, with and without including the effect of SFR as a second parameter.

\section{Summary}\label{sec:sec5}

We study the correlation between PAH and CO(1--0) emission in galaxies, with the aim of elucidating the underlying physical connection between these two components of the interstellar medium and to explore the possibility of using PAH emission as a proxy for estimating molecular gas content. We use the method of \cite{Zhang et al. 2021} to extract robust, spatially resolved measurements of the total ($5-20\ \mum$) PAH emission by combining mapping-mode Spitzer/IRS observations with mid-IR images. We focus on the subset of 19 nearby galaxies that overlap between the SINGS and BIMA SONG projects, the former providing the Spitzer MIR spectroscopy and ancillary optical-MIR images necessary to measure PAH fluxes and SFRs, and the latter interferometric CO(1--0) maps of comparable resolution to derive molecular gas masses. H~{\small I} maps of similar resolution from the THINGS collaboration are available for 13 of the 19 galaxies. The sample contains a mixture of star-forming galaxies and low-luminosity AGNs. The observations map the central $\sim$30\arcsec$\times$50\arcsec\ region of each galaxy, except for M\,51, which has a larger coverage of $\sim$40\arcsec$\times$240\arcsec, at a spatial resolution of 10\arcsec, which corresponds to a physical scale of $\sim 0.2 - 0.7$~kpc.

We confirm that PAH emission is tightly correlated with CO emission on sub-kpc scales, for both star-forming and active galaxies, as well as for different environments across a single galaxy. The same does not hold for H~{\small I} gas. We argue that the correlation between CO and PAH emission is not an artifact of their mutual relation to SFR. Instead, we suggest that the CO-PAH correlation reflects an underlying physical connection between PAHs and the diffuse interstellar gas, likely tied to a common heating mechanism. We present an empirical calibration to estimate molecular gas mass from the luminosity of PAH emission that has a total scatter of only $\sim 0.2-0.25$ dex. Including the SFR as a second parameter in principle can improve the accuracy of the molecular gas mass prediction, but the current sample is too limited to yield a firm conclusion on the merits of this approach. Calibrations using photometric estimates of PAH strength based on the Spitzer/IRAC4 and WISE/W3 filters give surprisingly similar results.

\acknowledgments
We thank the anonymous referee for constructive suggestions. This work was supported by the National Science Foundation of China (11721303, 11991052, 12011540375) and the China Manned Space Project (CMS-CSST-2021-A04, CMS-CSST-2021-A06).

\appendix
\section{Results Based on PAH Strength Derived from MIR Photometry}
\label{sec:AppA}

\begin{figure*}
\figurenum{A1}
\center{\includegraphics[width=1\linewidth]{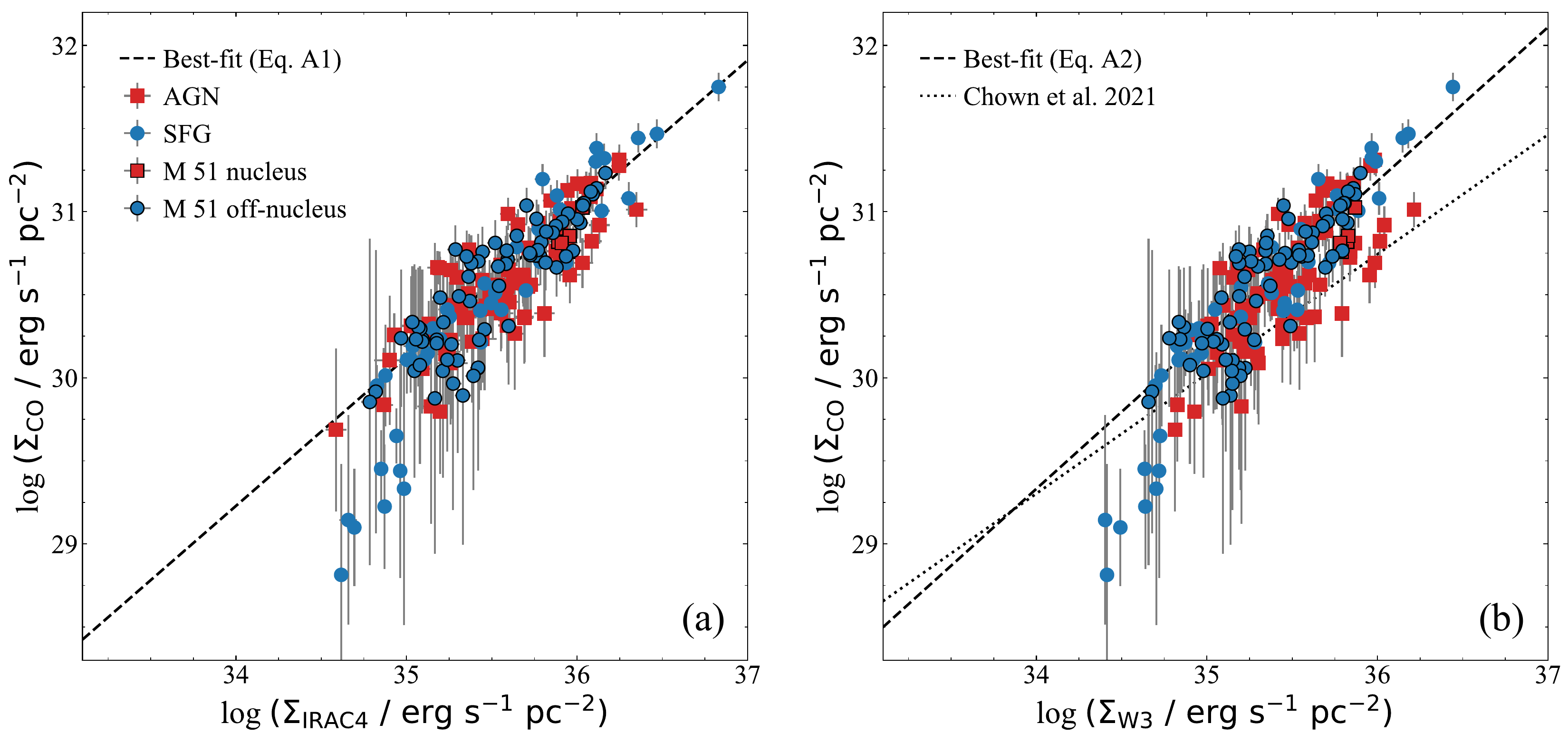}}
\caption{{The correlation between the surface brightness of CO(1--0) emission and emission in the MIR (a) $\rm IRAC4$ and (b) W3 bands.  Red squares and blue circles correspond to spaxels from the central regions of AGNs and SFGs, respectively, while the corresponding symbols with black borders denote the nuclear (AGN-like) and off-nuclear (SFG-like) spaxels from M\,51. The dashed lines indicate the best-fit linear regressions (Eqs.~\ref{equ:IRAC4_CO} and \ref{equ:W3_CO}), while the dotted line is taken from Figure~5b of \cite{Chown et al. 2021}. 
}\label{fig:MIR_CO}}
\end{figure*}

As has been investigated extensively, some broad-band MIR filters (e.g., Spitzer/IRAC4: \citealt{Regan et al. 2006}, \citealt{Bendo et al. 2010}; WISE/W3: \citealt{Gao et al. 2019}, \citealt{Chown et al. 2021}) are sensitive to the most prominent PAH features, offering a highly efficient alternative method to estimating PAH emission. These methods are attractive, in light of their applicability to large data sets available from large-area sky surveys. Figure~\ref{fig:MIR_CO} confirms that within our subsample of spatially resolved SINGS galaxies, the surface brightness of CO emission correlates strongly with the surface brightness measured in the IRAC4 and W3 bands. The best-fit regressions (Equations~\ref{equ:IRAC4_CO} and \ref{equ:W3_CO}) have very similar slope and scatter compared to those obtained using spectroscopically derived PAH measurements (Figure~\ref{fig:PAH_Gas}a; Equation~\ref{equ:PAH_CO}). Here, we define $L_{\rm band} = \nu L_{\nu}({\rm band})$, and we follow the prescription of \cite{Helou et al. 2004} to correct the IRAC4 band for stellar emission based on the IRAC1 map, with a scaling factor of 0.232. As a comparison, Figure~\ref{fig:MIR_CO}b also presents the pixel-based calibration between the surface brightness of CO emission and that in the W3 band, as presented by \cite{Chown et al. 2021}. Note that the calibration of \cite{Chown et al. 2021} is based on measurements of relatively lower surface brightness, as most of their resolved pixels are from the outer regions of galaxies.

\begin{equation}\label{equ:IRAC4_CO}
\log\,\Sigma_{\rm CO} = (0.894\pm0.035)(\log\, \Sigma_{\rm IRAC4} - 34.0) + (29.229\pm0.061);
\ \ \ \ \epsilon_t =  0.236\,\rm dex
\ \ \ \ \epsilon_i =  0.100\,\rm dex.
\end{equation}

\begin{equation}\label{equ:W3_CO}
\log\,\Sigma_{\rm CO} = (0.926\pm0.039)(\log\, \Sigma_{\rm W3} - 34.0) + (29.332\pm0.061);\ \ \ \ \ 
\ \ \ \ \epsilon_t =  0.250\,\rm dex
\ \ \ \ \epsilon_i =  0.128\,\rm dex.
\end{equation}

%\newpage

\end{document}